\documentclass[12pt]{article}
\usepackage{epsfig}

\usepackage{amssymb}
\usepackage{amsfonts}

\usepackage{color}
 
\def\be{\begin{equation}}
\def\ee{\end{equation}}
\def\ba{\begin{array}{c}}
\def\ea{\end{array}}

\newcommand{\bea}{\begin{eqnarray}}
\newcommand{\eea}{\end{eqnarray}}

\newcommand{\pbr}{\prec\!\!}
\newcommand{\pkt}{\!\!\succ\,\,}
\newcommand{\kt}{\rangle}
\newcommand{\br}{\langle}

\begin{document}

\begin{center}

{\Large \bf

Phase transitions in quasi-Hermitian quantum models
at exceptional points of order four

}

\end{center}

\vspace{0.4cm}

\begin{center}

  {\bf Miloslav Znojil}$^{a,b,c,d,}$\footnote{%corresponding author,
{e-mail: znojil@ujf.cas.cz}}

\end{center}

\noindent
$^{a}$The Czech Academy of Sciences,
 Nuclear Physics Institute,
 \v{R}e\v{z} 292,
250 68 Husinec, Czech Repubic

\noindent
$^{b}$Department of Physics, Faculty of
Science, University of Hradec Kr\'{a}lov\'{e}, Rokitansk\'{e}ho 62,
50003 Hradec Kr\'{a}lov\'{e},
 Czech Repubic

\noindent
$^{c}$%Research Fellow
School for Data Science and Computational Thinking, Stellenbosch
University, 7600 Stellenbosch,
 South Africa

\noindent
$^{d}$Institute of System Science, Durban University of
Technology, 4000 Durban,
%{4001}
 South Africa

% \vspace{10mm}
 %\newpage

%\textcolor{black}{...}

%%%%%%%%%%%%%%%%%%%%%%%%%%%%%%%%%%%%%%%%%%%%%%%%%%%%%%%%%%%%%%%%%%%%
\subsection*{Abstract}

Phase transition in quantum mechanics is interpreted
as an evolution, at the end of which,
typically, a parameter-dependent
and Hermitizable Hamiltonian $H(g)$
loses its observability. In the language of mathematics,
such a ``quantum catastrophe'' occurs at an
exceptional point
of order $N$ (EPN).
Although
the Hamiltonian $H(g)$ itself
becomes unphysical in the limit of
$g \to g^{EPN}$,
it is shown that it
can play the role of
an unperturbed operator
in an innovative perturbation-approximation analysis of the
vicinity of the EPN singularity.
As long as such an analysis
is elementary at  $N\leq 3$
and purely numerical at $N\geq 5$,
we pick up $N=4$ and demonstrate
that for an arbitrary quantum system,
the specific
(i.e., already sufficiently phenomenologically rich)
EP4 degeneracy
becomes accessible
via a unitary evolution process.
This process is shown
realizable inside a parametric domain ${\cal D}_{\rm physical}$,
the boundaries
of which are determined, near $g^{EP4}$, non-numerically.
\textcolor{black}{Possible relevance of such a mathematical result
in the context of non-Hermitian photonics is emphasized.}

%\newpage

\subsection*{Keywords}.

quantum mechanics using quasi-Hermitian observables;

non-Hermitian but Hermitizable {\it alias\,} quasi-Hermitian Hamiltonians;

\textcolor{black}{unitary} quantum phase transitions at exceptional points;

specific fourth-order exceptional points (EP4);

\textcolor{black}{relevance of non-numerical study of EP4s in photonics};

\newpage

\section{Introduction}

\textcolor{black}{At present, a deep and productive relevance
of an originally purely mathematical concept of
exceptional point (EP, \cite{Kato})
in photonics is well known \cite{Rotter}.
From a historical perspective
(cf., e.g., the introductory chapter in
dedicated monograph \cite{book}),
one could really localize several independent roots of
the related successful theoretical as well as experimental
developments.}

\textcolor{black}{During their first stage, for example,
a key driving force originated from
the relativistic quantum field theory, in the framework of which
the challenges emerged due to the discoveries of divergences
of many innocent-looking perturbation expansions.
Soon, these phenomena
found some of their very natural
though not quite expected mathematical explanations
precisely
in the existence of the EP
values of certain relevant
phenomenological parameters (cf., e.g., \cite{BW,Alvarez}).}

\textcolor{black}{The currently popular idea
of the use of the EPs
in photonics
only emerged
when several authors realized
that there must exist an intimate connection between
the Schr\"{o}dinger equation
of quantum mechanics
(in which the EPs already found their applications)
and the classical Maxwell equations
(especially when considered in
the so-called paraxial approximation,
cf., e.g., \cite{Mousse}).
In any case, the subsequent progress
was truly impressive (cf., e.g.,
the recent review \cite{Christodoulides}),
based mainly on a wide range of
mathematics
shared by the quantum theory and photonics.
Incidentally, this also motivated our present
paper in which we are going to restrict our attention
to several rather interesting
interdisciplinary aspects
of the shared mathematics
using, in most cases, just the
representative language of
the so-called quasi-Hermitian quantum models
in which the EPs manifest themselves as the
instants of a specific class of
the so-called quantum phase transitions.}

The task of description of
a quantum phase transition is,
for several reasons, difficult. First of all,
one has to leave the comparably comfortable
formalism of
classical mechanics.
After the quantization, the
numbers representing the
classical observable
quantities like, e.g.,
a
(possibly, time-dependent)
point-particle momentum $p=p(t)$
must be replaced by operators.
In particular, {the quantized energy}
$E$, conserved or not, {has to be treated as an eigenvalue} of
a preselected energy-representing Hamiltonian $H$, etc.
Thus, a more or less purely geometric nature of the
classical phase-transition theories (cf., e.g., \cite{Zeeman}) is lost.

Secondly, all of the operators $\Lambda(t)$ representing a relevant
quantum observable
have to be diagonalizable and, moreover,
they are usually chosen or constructed as
self-adjoint in
a suitable (and, mostly, infinite-dimensional)
Hilbert space of states ${\cal H}_{\rm physical}$.
In such a traditional quantum model-building setup,
the very existence of a genuine ``change of phase''
(i.e., basically, of a loss of the reality of
the spectrum of at least one of the observables)
seems to contradict the well known tendency of
eigenvalues of any self-adjoint operator
to avoid any phase-transition-mimicking merger.
Just an apparent ``repulsion of levels'' is
encountered in many experiments.

On the level of abstract quantum theory,
both of the latter obstacles
can be circumvented when one resorts
to the
less usual formulation
of quantum mechanics
called quasi-Hermitian quantum mechanics (QHQM,
cf. one of its oldest reviews
\cite{Geyer} for a comprehensive introduction).
The resolution of at least some
of the quantum phase-transition puzzles
is offered there by the possibility
of the realization
of the phase-transition-mimicking mergers
of the eigenvalues as well as of the states
at a singularity
called, by mathematicians,
exceptional point (EP,  \cite{Kato}).
By definition, this means that
in the singular,
manifestly unphysical EP limit,
the overall spectrum
of an observable under consideration
remains real but, at the same time,
it degenerates and
ceases to admit standard
interpretation.
Such a
way of thinking about quantum phase-transition
processes
will be also accepted in our present paper.

We will start from the observation that
in the majority of the existing
constructions of models admitting an EP-related
phase transition,
the authors
prefer
the
systems characterized by the
observability of the mere
parameter-dependent bound-state energy levels $E_n(g)$ with $n=0,1,
\ldots\,$.
For the sake of definiteness,
moreover,
just the models admitting the most elementary
EP mergers
of order two (EP2) are usually considered.
In the language of mathematics this means that
the ``first nontrivial'' form of the
dynamical scenario is only being described, in which
the loss of
the diagonalizability of the  Hamiltonian $H(g)$
in the
limit of $g\to g^{(EP2)}$
is restricted to a two-dimensional subspace
of the whole Hilbert space.
In other words,
the EP2-related merger of energies $E_{n_j}(g)$
as well as of the related bound-state eigenvectors $|\psi_{n_j}(g)\kt$
only involves a single pair of states with subscripts $j=1$ and $j=2$.

Even the EP2-related studies need not be trivial.
In the non-Hermitian physics framework
admitting complex energies (i.e., resonances -- cf., e.g.,
an introductory outline
of the underlying theory in monograph \cite{Nimrod}),
one of the
oldest and, at the same time, one of the
best known illustrative examples of the EP2 mergers
has been
found, many years ago,
by Bender with Wu \cite{BW}.
In their methodically motivated considerations concerning
the general relativistic quantum field theory
they paid attention just to a zero-dimensional version of this theory.
More specifically, they considered
the quantum-mechanical
anharmonic-oscillator Hamiltonian $H(g)=p^2+x^2+g\,x^4$
with $x \in (-\infty,\infty)$,
and they
managed to localize
its EP singularities. All of them appeared to lie
off the real axis,
i.e., at the complex values of $g= g^{(EP2)}_{m,n}$
numbered by the two integers $m,n=0,1,\ldots$.
Although all of them belonged just to the simplest,
second-order category, their discovery was important:
These degeneracies
became known
as the
Bender-Wu singularities \cite{Alvarez}.

Out of quantum physics,
some authors also speak about
non-Hermitian degeneracies~\cite{Berry,Heiss,Heissb}.
Beyond all of their most elementary EP=EP2
models, quantum or non-quantum
(see also their most recent
samples in \cite{SciRep,Nimrode,Bijan}), there appeared
studies of
certain more complicated, viz.,  EP3-supporting
phase-transition models \cite{ep3,ep3b}.
The point was that the authors
of the latter generalizations
(in which the degeneracy involves a triplet of states)
could make use of the
exact solvability of the related secular equations
(i.e., of the mere algebraic polynomial equations
of the third order in the energy)
in terms of the well known Cardano formulae.

The existence of such a form of solvability also inspired
our present paper.
We imagined that
the next, EP4-supporting
family of models is in fact the only remaining
class of the special quantum systems
enjoying the advantage of being non-numerical.
Indeed, the related algebraic polynomial
secular equations (of the fourth order) are
in fact the last ones
solvable in closed form.
Unfortunately, in contrast to Cardano formulae,
the closed formulae for the quartic-polynomial roots
can hardly be considered user-friendly.
For this reason, the EP4-supporting
quantum phase-transition models
have only marginally and not too deeply been considered
in the literature \cite{corridors}.

In our present paper we decided to fill the gap.
The presentation of our results will be
preceded by section \ref{pragma} in which
we will outline, briefly, an overall correspondence between
the exceptional points and the quantum phase transitions.
Next, in
section \ref{theora} we will
restrict our attention
to the case of the so-called closed
quantum systems in which, by definition,
the evolution remains unitary.
We will explain there
that the quantum phase transitions
may exist
even under such a
constraint,
provided only that
one
makes use of the
above-mentioned QHQM
description of the dynamics.
For the sake of a maximal completeness
of the paper, also
a few basic features of the QHQM reformulation
of the standard quantum theory
are there added.

Our main message
will finally be presented
in sections \ref{dogma} and \ref{segma}
in which we will formulate the problem
and describe its solution.
The basic idea
of our approach
will lie in a reduction of the
general
EP4-related phase-transition problem
to its analysis in a small vicinity
of the EP4 degeneracy.
This will enable us to
construct its sub-vicinity
${\cal D}_{\rm physical}$
(i.e., a  unitarity-compatible subdomain
of the full space of the
available dynamical parameters)
and to localize its boundaries
using a suitable form of mathematical
perturbation theory.

These results are then discussed
and summarized in section \ref{discussion}.

%\newpage

\section{Exceptional points\label{pragma}}

\subsection{Unitary- and non-unitary-evolution scenarios}

In some applications of quantum mechanics
it need not be easy to distinguish between
the systems characterized by the
unitary and non-unitary forms of
evolution.
During the study of the former scenarios
(with the models describing the so-called
closed quantum systems),
the main emphasis is
usually put on a completeness and on a maximal theoretical
consistency of the picture \cite{Messiah}.
In contrast,
the admissibility of
a manifestly non-unitary dynamics
finds its strongest support
in experimental physics \cite{Ingrid,SSHb}.
One feels more inclined there to admit
that
we only rarely manage to separate the system
from its ubiquitous but only partially
understood  environment.
Thus, it makes sense to speak about the
open quantum systems in such a case
\cite{Nimrod}.

In both of the latter,
phenomenologically non-equivalent
situations one can
encounter the phenomena
in which
it is not even too easy to
separate
the unitary and non-unitary aspects of
the evolution.
As an example we could recall the
models in which some, but not all, of the operators
representing an observable (i.e., say,
a parameter-dependent Hamiltonian $H(g)$)
will lose their diagonalizability and, hence,
the status of a quantum observable.
For an explicit illustration of such a rather involved scenario
the authors of review paper \cite{Geyer}
recalled the traditional two-parametric
Lipkin-Meshkov-Glick
multiparticle model
in a
generalized form characterized by
a non-Hermitian Hamiltonian operator
(cf. equation Nr.~(3.1) in {\it loc. cit.}).
Manifestly these authors demonstrated that their model
really admitted several rather realistic forms of a
genuine quantum phase transition.

A few basic qualitative
features of the example may be found summarized
in the phase diagram Nr. 1 of {\it loc. cit}.
In this diagram, indeed, only some subdomains
of the space of the parameters appeared to admit a closed-system
interpretation of the dynamics of the system,
with some of their boundaries being,
precisely, the EP singularities mimicking
certain rather realistic forms of the quantum
phase transitions.

\subsection{Non-Hermitian degeneracies}

In a more general, model-independent manner, by definition,
the EP singularities
will be,
in general,
of order $N$
(EPN, \cite{Kato}).
At $g \neq g^{(EPN)}$,
before we move to the EPN
extreme and to the phenomenon of the
$N-$tuple degeneracy,
we may assume that
this degeneracy is going to involve
an $N-$plet of  non-degenerate energy-representing
eigenvalues $E_n(g)$ as well as, simultaneously,
the related eigenstates $|\psi_n(g)\kt$
of the Hamiltonian $H(g)$
with
$n=n_1,n_2,\ldots,n_N$.
Once we assume that
both of these $N-$plets happen
to merge in the EPN limit, we may write
 \be
 \lim_{g \to g^{(EPN)}}\,E_{n_j}(g)= {E^{(EPN)}}\,,
 \ \ \ \
 \lim_{g \to g^{(EPN)}}\,|\psi_{n_j}(g)\kt= {|\psi^{(EPN)}\kt}\,,
 \ \ \ \
 j=1,2,\ldots,N\,.
 \ee
Under certain
standard mathematical assumptions \cite{Kato}
there will also exist a vicinity
of the critical parameter
such that
$g \neq g^{(EPN)}$, inside which
the operator $H(g)$ itself
remains diagonalizable and tractable
as a quantum Hamiltonian.
In the limit
$g \to g^{(EPN)}$
(and after the restriction of our interest
to the mere relevant, $N-$dimensional subspace of
the Hilbert space), in contrast,
the conventional diagonalizability,
i.e., the solvability of
the conventional
Schr\"{o}dinger equation
 \be
 H^{(N)}(g) |\psi_n(g)\kt = E_n(g)\,|\psi_n(g)\kt
 \,,\ \ \ \ n = 0, 1, \ldots,N
 \label{cose}
 \ee
becomes replaced, in
the same subspace,
by the solvability
of another, unphysical $N$-by-$N$-matrix
equation
 \be
{H}^{(N)}(g^{(EPN)})
\,
{\cal U}^{}
={\cal U}^{}
\,
J^{(N)}(E^{(EPN)})\,.
 \label{debyeq}
 \ee
Its solution
${\cal U}={\cal U}^{(EPN)}$
is called transition matrix, while
the symbol
 $$
 J^{(N)}(x)=
 \left[ \begin {array}{ccccc}
 x&1&0&\ldots&0
 \\
 0&x&1&\ddots&\vdots
 \\
 0&0&\ddots&\ddots&0
 \\
 \vdots&\ddots&\ddots&x&1
 \\
 0&\ldots&0&0&x
 \end {array} \right]
 $$
stands for the so-called
Jordan-matrix function of the
degenerate, unphysical energy $x=E^{(EPN)}$.

\subsection{Phase transitions}

The
construction of the $N$-by-$N$-matrix
solution ${\cal U}^{(EPN)}$ of the
canonical,
Schr\"{o}dinger-equation-resembling Eq.~(\ref{debyeq})
is important in applications.
It opens, in particular, the way towards a
decisive simplification of
the matrix form of every physical Hamiltonian $H(g)$
in which the parameter $g$ is different
from $g^{(EPN)}$
but still
lying not too far from this critical value.
Thus, for the sake of definiteness,
we assume that the
absolute value of the difference $\lambda=g-g^{(EPN)}$
remains small.
In such a situation we
have to reconsider Schr\"{o}dinger
Eq.~(\ref{cose}), treating it
as a $\lambda-$dependent eigenvalue problem
 \be
 H^{(N)}(g^{(EPN)}+\lambda) |\psi_n(g^{(EPN)}+\lambda)\kt
 = E_n(g^{(EPN)}+\lambda)\,|\psi_n(g^{(EPN)}+\lambda)\kt
 \,,\ \ \ \ n = 0, 1, \ldots,N\,
 \label{ucose}
 \ee
i.e., as a rather unusual but mathematically well motivated
perturbation of its manifestly unphysical EPN limit.

At a sufficiently small $\lambda$, we now have to use
perturbation theory, say,
in its QHQM-adapted form of Ref.~\cite{pertsym}, proceeding
along the lines as described, in detail, e.g.,
in papers \cite{admissible,pertEP}.
Naturally, even when we
restrict the range of the parameter $\lambda$
to a small vicinity of zero, we can still
speak about
the
phase-transition
process controlled by
the variable
$g$
reaching its  loss-of-the-observability extreme in the EPN limit
of $g \to g^{(EPN)}$.

During the implementation of perturbation theory
we reveal that
our knowledge of
the
solutions ${\cal U}={\cal U}^{(EPN)}$
of the unperturbed and
$g-$independent Schr\"{o}dinger Eq.~(\ref{debyeq})
proves particularly useful because
it
enables us to
simplify
the  perturbation-approximation
construction via a
replacement of
any initial perturbed, $g-$dependent Hamiltonian $H(g)$
by its isospectral $\lambda-$dependent avatar
 \be
  P(\lambda)={\cal U}^{-1}\,H(g^{(EPN)}+\lambda)\,
 {\cal U} =J^{(N)}(E^{(EPN)})+
 {\rm a\ perturbation}\,.
 \label{pertie}
 \ee
The point is that
at the sufficiently small $\lambda$s,
the new matrix $P(\lambda)$
remains
dominated by its unperturbed Jordan-matrix component
so that the systematic construction of the
perturbed bound-state
solutions
becomes perceivably facilitated.

%\newpage

\section{Closed systems admitting phase transitions\label{theora}}

We saw above that
the description of
the phase transition, i.e., of
the passage of the quantum system in question
through its EPN singularity may acquire an explicit
perturbation-approximation form when we
employ a
parameter
$\lambda$ which would be properly time-dependent.
Marginally, let us add that
the perturbed energy spectra near EPNs are, in general,
complex.
In the literature, nevertheless, their study found a persuasive
mathematical motivation, first of all, in the abstract perturbation
theory \cite{Kato}. Soon, nevertheless, the EPN-related mathematics
found also important applications in physics.
Nevertheless,
as long as all of the
energies near an EP
singularity are, in general, complex,
one must interpret the
corresponding
states as resonances, with the
non-Hermitian system in question
acquiring the status of the so-called open quantum system.

In the case of the other, unitary, closed quantum systems
(for which the states are stable of course),
one has to
add a few new conditions to
guarantee
the reality of the
energies.
This is a
fairly complicated
task,
a fulfillment of which will be discussed
in the rest of this paper.

\subsection{Probabilistic interpretation of quasi-Hermitian observables}

In the majority of textbooks on
quantum mechanics
the authors pay attention, first of all,
to the
unitary-evolution processes
and to their rather detailed
description
in a specific representation
called Schr\"{o}dinger picture \cite{Messiah}.
In such a
very traditional setting, the time-evolution
of the
state-representing
wave functions $|\psi\pkt$
in a suitable physical Hilbert space
(say, ${\cal L}$)
has to be generated by a conventional
diagonalizable and
self-adjoint
Hamiltonian operator (say, $\mathfrak{h}=\mathfrak{h}^\dagger$).

A transition from the latter, conventional
formulation of quantum theory
to the non-Hermitian QHQM
formalism
can be traced back to
the papers by Dyson \cite{Dyson}.
Emirically, he revealed
that
for the variational-calculation purposes,
the conventional textbook form $\mathfrak{h}$
of the Hamiltonian need not be optimal, and that
a truly significant enhancement of the efficiency
of the calculations can be achieved via
a suitable isospectral
but non-unitary preconditioning
of this operator,
 \be
 \mathfrak{h}\ \to  \
 H=\Omega^{-1}\,
 \mathfrak{h}\,
 \Omega\,,\ \ \ \ \ \Omega^\dagger\Omega=\Theta \neq I\,.
 \label{zasedys}
 \ee
From the point of view of the
prediction of the
results of experimental measurements,
both of the operators $\mathfrak{h}$ and $H$
describe the same
bound states
(i.e., the same closed and unitary quantum system).
Thus, the Dyson-proposed preference of the use of $H$
is to be understood as just an {\it a posteriori\,}
observation of a purely technical aspect of his specific model.

In a fully abstract theoretical setting,
the mapping (\ref{zasedys}) can be inverted.
Still,
one can easily verify that the Hermiticity
of $\mathfrak{h}$ in ${\cal L}$
is formally equivalent, for non-Hermitian $H\neq H^\dagger$,
to its quasi-Hermiticity property
 \be
 H^\dagger\,\Theta=\Theta\,H\,
 \label{[7]}
 \ee
{\it alias\,} $\Theta-$quasi-Hermiticity property valid
in the corresponding
Hilbert space,
say, ${\cal H}_{\rm mathematical}$.

Admissibly \cite{Dyson,SIGMA}, the latter space need not coincide with
${\cal L}$ so that,
due to the non-Hermiticity of $H$, the
space ${\cal H}_{\rm mathematical}$
itself is unphysical, i.e., it is
not equivalent to the above-mentioned
QHQM Hilbert space ${\cal H}_{\rm physical}$.
Fortunately,
both of these spaces can be chosen as sharing
their ket-vector elements $|\psi\kt$.
Then, the only difference
involves the respective inner products.
Once we accept the conventional bra-ket denotation
$\br \psi_1|\psi_2\kt$ of the inner product
in ${\cal H}_{\rm mathematical}$,
we may succeed in representing ${\cal H}_{\rm physical}$
in the same (but, by assumption, user-friendlier)
representation space ${\cal H}_{\rm mathematical}$.
For this purpose it is sufficient to
change the inner product and use
the following metric-mediated ``correct
physical product'' overlap
 \be
 \br \psi_1|\Theta |\psi_2\kt\,.
 \ee
The latter values, incidentally, coincide with the
conventional textbook inner products
$\pbr \psi_1|\psi_2\pkt$
in ${\cal L}$: To see this, it is sufficient to
recall Eq.~(\ref{zasedys})
and to conclude that $|\psi\pkt = \Omega\,|\psi\kt$
(cf., e.g., \cite{SIGMA} for the reasons and merits
of such a slightly generalized
version of the Dirac's notation convention).

For an efficient explicit
description of an EP-related phase transition
in a closed quantum system,
it may prove sufficient to work with
a suitable non-Hermitian
and parameter-dependent Hamiltonian operator
$H=H(g)$ with real spectrum
admitting an EPN singularity at $g=g^{(EPN)}$.
One only has to keep in mind that
the existence of the Dyson-map
correspondence (\ref{zasedys})
as well as of the
related correct physical inner-product metric $\Theta$
is ``fragile'' \cite{fragile,catast}.
In the limit of $g \to g^{(EPN)}$,
by definition, both of these operators
would have to cease to exist.

In order to make the theory internally consistent and complete,
our task becomes twofold.
For a given $H(g)$
we have to demonstrate, firstly, the existence of the
EPN singularity. Secondly,
we have to demonstrate
the existence of an unfolding of this
singularity during which the spectrum remains all real.
Both of these tasks are fairly nontrivial and
requiring, in general, a
purely numerical treatment.

\subsection{Exactly solvable benchmark models}

For the closed quantum systems
admitting the second-order (i.e., $N=2$) and
the third-order (i.e., $N=3$)
non-Hermitian EPN degeneracies,
the realization of both of the two above-mentioned
tasks (viz., of the EPN localization and
of the specification of a path of a unitary unfolding
of the singularity)
can already be found in the literature.
In fact, the true difficulties only occur at $N=4$
which is still solvable exactly and non-numerically,
in principle at least.
Beyond this case, all of the $N\geq 5$ EPN models
can only be treated,
on a suitable approximation level, numerically,

Even at the arbitrarily large $N$s,
there are
solvable-model
exceptions od course \cite{catast}.
In Ref.~\cite{passage}, for example,
the readers may find a commented reference to a
certain  fairly realistic
many-body Bose-Hubbard (BH)
quantum system as proposed
and studied,
in a broader
phenomenological context, by Graefe et al \cite{Uwe}.
Out of the sequence
of the underlying
$N$ by $N$ Hamiltonian matrices
of a variable dimension $N$
we picked up, in {\it loc. cit.},
the strictly quasi-Hermitian subset
  \be
 H^{(2)}_{(BH)}(g)=\left[ \begin {array}{cc} -ig&1
 \\\noalign{\medskip}1&ig\end {array} \right]
 \,,\ \ \ \ \
% Q^{(2)}_{(BH)}=\left[ \begin {array}{cc} -i&1
%\\\noalign{\medskip}1&0\end {array}
% \right]\,,
 %$$
%%We also get the sufficiently elementary next pair,
% $$
 H^{(3)}_{(BH)}(g)=\left[ \begin {array}{ccc}
 -2\,ig&\sqrt {2}&0\\\noalign{\medskip}\sqrt {2}&0&\sqrt {2}
 \\\noalign{\medskip}0&\sqrt {2}&2\,ig\end {array}
 \right]
 \,,\ \ \ \ \ldots
 %Q^{(3)}_{(BH)}= \left[ \begin {array}{ccc} -2&-2\,i&1
 %\\\noalign{\medskip}-2\,i\sqrt {
%2}&\sqrt {2}&0\\\noalign{\medskip}2&0&0\end {array} \right]\,,\
%\ldots\ .
 \label{speic}
 \ee
having real spectra if and only if
$|g|\leq 1$.
We could treat this system as
closed (i.e., unitary) at
$|g| < 1$.
Moreover, we showed, constructively, that
due to a certain immanent symmetry, the
specific models (\ref{speic})
happen to admit, at any $N$,
a quantum phase transition
in the two EPN limits of $g \to \pm 1$.
The choice of $|g| < 1$
forms then a
corridor of passage
from the standard unitary dynamical regime
to the corresponding,
phase-transition-representing EPN
singularity.

Among the most important features of the
exactly solvable quasi-Hermitian
quantum models as sampled
by Eq.~(\ref{speic}) one can count, first of all,
the existence of the
closed formulae for the real bound-state energies
-- in \cite{passage},  these formulae
were displayed and shown valid
at all of
the physical couplings $g < g^{(EPN)}$.
Another specific merit of the Bose-Hubbard-like
toy-model family of Eq.~(\ref{speic}) has to be seen in the
availability of the
exact $N$-by-$N$
transition-matrix solutions
of
the unperturbed Schr\"{o}dinger
Eq.~(\ref{debyeq}), say, at $g=g^{(EPN)}=1$,
  \be
 %H^{(2)}_{}(g)=\left[ \begin {array}{cc} -ig&1\\\noalign{\medskip}1&ig\end {array} \right]
% \,,\ \ \ \ \
 {\cal U}^{(2)}_{(BH)}=\left[ \begin {array}{cc} -i&1\\\noalign{\medskip}1&0\end {array}
 \right]\,,\ \ \ \ \
 %$$
%%We also get the sufficiently elementary next pair,
% $$
 %H^{(3)}_{}(g)=\left[ \begin {array}{ccc}
% -2\,ig&\sqrt {2}&0\\\noalign{\medskip}\sqrt {2}&0&\sqrt {2}
% \\\noalign{\medskip}0&\sqrt {2}&2\,ig\end {array}
% \right]
% \,,\ \ \ \ \ldots
 {\cal U}^{(3)}_{(BH)}= \left[ \begin {array}{ccc} -2&-2\,i&1\\\noalign{\medskip}-2\,i\sqrt {
2}&\sqrt {2}&0\\\noalign{\medskip}2&0&0\end {array} \right]\,,\ \ \ \
\ldots\,.
 \label{espeic}
 \ee
Indeed,
as we already emphasized above,
the knowledge of ${\cal U}$
in the ``unphysical'',
auxiliary EPN limit of $g \to g^{(EPN)}$
is a key to the conversion
(\ref{pertie})
of the ``realistic'', perturbed physical
Hamiltonian $H(g)$ with $g \neq g^{(EPN)}$
to its
representation $P(\lambda)$
which is more
user-friendly.

Even beyond the class of solvable models,
the representation $P(\lambda)$
of any preselected
perturbed Hamiltonian
is, in some sense, universal
and maximally economical
because
it allows us to ignore all of the
inessential parameters which
just remain encoded in ${\cal U}$,
and
which do not enter directly
the
simplified
canonical Hamiltonian $P(\lambda)$.

%\newpage

\section{Fourth-order exceptional points\label{dogma}}

A realization of the fall
of a quantum system
into its EPN
singularity
can be studied
via Schr\"{o}dinger  Eq.~(\ref{ucose}) at a small $\lambda$.
A basic technical
tool of its perturbation-expansion solution
lies then in the specification of unperturbed basis,
the role of which is to be played by
the $N$ by $N$ matrix solution ${\cal U}$
of a rather anomalous unperturbed Schr\"{o}dinger Eq.~(\ref{debyeq}).

For our
illustrative sequence of Hamiltonian matrices (\ref{speic}),
in particular,
matrices ${\cal U}^{(N)}$
were obtained, in \cite{passage},
in the form
of sequence (\ref{espeic}),
with its individual
$N$-by-$N$-array elements
defined, at all $N$,
in terms of certain compact algebraic
(i.e., full-precision) expressions.
Now, we are going to turn attention
to the more general classes of, potentially, more
realistic quantum models admitting, in general, just
an approximative treatment.

\subsection{The choice of $N=4$}

All of the solvable-model results
as mentioned in section \ref{theora}
enhanced
our interest in the
possibility of an analysis
of the EPN-generated quantum phase transitions
in a model-independent
but still analytic and
manifestly non-numerical manner.
This led us immediately to the
choice of $N=4$.
Obviously, besides the related, purely mathematical
challenge given by its last-solvable-model status,
we also imagined that the older,
physics-motivated studies of
the quasi-Hermitian quantum systems
exhibiting the EPN non-Hermitian degeneracies
with $N \leq 3$
still could not cover
a wealth of possible phenomena emerging at the larger $N$s.

Naturally, all of the theoretical as well as experimental physicists
active in the field
were very well aware of the fact that
the costs of a move beyond the ``natural boundary'' of
$N = 3$ could be high.
In our eyes, nevertheless,
such a move
appeared to be really very well motivated by its
undeniable relevance in the applied quantum physics
concerning both of its open- and closed-system subcategories.
Moreover, we found a partial resolution of the
conflict between the appeal and feasibility
in the fact that
the mathematics needed in the
former, open-system subcategory remains elementary.
Thus, we decided to
reduce the task and pay attention to the
study of the mere EPN-related closed systems with $N=4$.

Marginally, we also felt encouraged by the
rather formal observation that
the EP4-supporting quantum systems offer and form, in fact, a unique
``missing link'' between the algebraic and numerical
tractability of the non-Hermitian but potentially unitary
phase-transition
processes.
In our present paper we intend to show, therefore, that
their realization remains analytic and exact
not only
for the
well known simplest scenarios with the
EPN orders $N=2$ and $N=3$,
but also in
the more realistic
but also more sophisticated and
less frequently studied systems with $N=4$.

\subsection{Six-parametric canonical Hamiltonian $P^{(N)}(\lambda)$ at $N=4$
\label{stigma}}

The choice of $N=4$ is special since
the
generic
input Hamiltonian $H^{(4)}(g)$ (i.e., not an exactly solvable
one) will still
be assigned just a
purely numerical form of ${\cal U}$ in applications.
The reason is that
in spite of the existence of the closed formulae,
their
use remains utterly impractical due to
their overcomplicated structure.
Nevertheless,
the possibility of having just an approximate,
numerical form of the transition
matrix ${\cal U}$ still
opens the way towards an efficient replacement of the
perturbed Schr\"{o}dinger Eq.~(\ref{cose}) or~(\ref{ucose})
by a simplified, much more user-friendly
alternative.

Beyond the very special class of the exactly solvable models,
the construction of
${\cal U}={\cal U}^{(N)}$
becomes numerical.
Even then,
there remain good reasons for
transition (\ref{pertie}) from
any preselected phenomenological input Hamiltonian
$H^{(N)}(g)$
to its
isospectral
version $P^{(N)}(\lambda)$.
Especially at the larger $N\geq 5$,
the diagonalization of $H^{(N)}(g)$
becomes replaced by
the diagonalization  of $P^{(N)}(\lambda)$
which can be performed, near the EPN singularity, perturbatively.

In
the $P(\lambda) \leftrightarrow H(g)$ correspondence
of
Eq.~(\ref{pertie}) with $N \geq 5$,
one must really start working with
the numerical form of the transition matrices.
In practice, the use of their numerical form
could prove recommendable even at $N=4$, in spite of the fact that
in this case,
the closed formulae also do exist.
One can make a choice
between the exact and approximate ${\cal U}^{(N)}$
but, in both cases, one has to guarantee that
the spectrum of the resulting simplified
perturbed Schr\"{o}dinger equation
 \be
 P^{(N)}(\lambda) |\phi_n(\lambda)\kt
 = E_n(g^{(EP4)}+\lambda)\,|\phi_n(\lambda)\kt
 \,,\ \ \ \ n = 0, 1, 2,\ldots,N\,
 \label{pucose}
 \ee
remains real,
especially
in a small vicinity of EPN, i.e.,
in the vanishing-perturbation limit of $\lambda \to 0$.
Thus,
the proofs of these properties
form one of our key mathematical
tasks since the related guarantee of the unitarity of evolution
forms the very core of the description of the
related quantum phase transition.

Whenever we decide to study
the phenomenon in
a closed quantum system which is
characterized by its $N$-by-$N$ matrix Hamiltonian $H^{(N)}(g)$
possessing, in a physical domain of
$g \in {\cal D}$, a real and non-degenerate spectrum
of energies $\{E_n(g)\}$,
we have to guarantee, first of all, that this model
really does exhibit an EPN
degeneracy in the limit of  $g \to g^{(EPN)}$.
Secondly, it also makes sense to require that
the value of $g^{(EPN)}$
is an element of the boundary
of the physical domain of parameters,
i.e.,
$g^{(EPN)}\in \partial (\cal D)$.
Under these assumptions
we may always use the standard tools of linear algebra and
construct the ``unperturbed-basis''
solution ${\cal U}^{(N)}$
of Eq.~(\ref{debyeq}).

As a consequence, we will be able to perform
the transformation (\ref{pertie}) yielding
our perturbed Schr\"{o}dinger equation~(\ref{ucose})
in its equivalent canonical form
(\ref{pucose}).
As long as we decided to restrict our attention
to the quantum systems which are quasi-Hermitian,
we have to keep in mind that
their evolution
which is non-unitary in the
mathematically preferable
but manifestly unphysical representation space
${\cal H}_{\rm mathematical}$
will have to be, simultaneously,
unitary in the ``correct space''
${\cal H}_{\rm physical}$.

In a way explained in Ref.~\cite{corridors},
this leads to the necessity of
a severe restriction imposed upon the class of the
admissible,
spectral-reality-supporting
perturbations in Eq.~(\ref{pertie}).
At $N=4$,
for an enhancement of the transparency of such a
procedure,
we will shift
the
origin on the real line of the bound-state energies,
set
$\eta=E-E^{(EP4)} $
and get $\eta^{(EP4)}=0$.
We
will also
shift the parameter $g$
yielding $g^{(EP4)}=0$. Finally, we will re-scale
the
measure $\lambda$ of the smallness of perturbations
(see \cite{pertEP} for the reasons).
As a consequence, our
Eq.~(\ref{pucose})
(i.e., the simplified
perturbed Schr\"{o}dinger equation
living in an $N-$dimensional EPN-related subspace
of ${\cal H}_{\rm mathematical}$
with $N=4$)
will read
 \be
 P^{(4)}(\lambda) |\phi_n(\lambda)\kt
 = \eta_n(\lambda)\,|\phi_n(\lambda)\kt
 \,,\ \ \ \ n = 0, 1, 2,3\,.
 \label{zucose}
 \ee
Here, Hamiltonian $P^{(4)}(\lambda)$
is still defined by a properly adapted
special case of Eq.~(\ref{pertie}) at
$N=4$. This means that such a matrix contains 16
matrix elements which carry a complete information about
the dynamics of the system in question at small $\lambda \neq 0$.

%\textcolor{black}{sem vloz (b)}

\textcolor{black}{At this stage of developments, several questions
have to be addressed in explicit manner.
The first
and the most important one concerns the phenomenological
and descriptive
role of
matrix $P^{(4)}(\lambda)$ in
experiments including, first of all,
non-Hermitian
(although, not necessarily, gain/loss) photonics.
Indeed,
such a matrix can only be perceived as
an optimal representation of dynamics
in a not too easily specified vicinity
of the
underlying EP4 degeneracy
of the system.}

\textcolor{black}{Even a brief
answer to such a question
would require
a return to Eq.~(\ref{pertie})
and, in particular,
to the original form of Hamiltonian $H(g^{(EP4)}+\lambda)$.
In it, expectedly, one still possesses a clear
and explicit knowledge
of the physical meaning of the parameter(s) $g$.
Only the isospectral one-to-one mapping $H(g^{(EP4)}+\lambda) \to
P^{(4)}(\lambda)$
enables us to
clarify the criteria of
the smallness of perturbation. At the same time,
the connection of the latter matrix with the
experimentally relevant parameters only remains encoded
in the purely formal,
optimal-basis-determining transition matrix ${\cal U}$.}

\textcolor{black}{The second important phenomenology-oriented question
is related to the above-emphasized distinction between
the real and non-real (i.e., partially or completely complex) spectra.
A return to the need of a better understanding of this
distinction would re-open not only the problem of
a consequent separation of the closed quantum
systems from the open quantum systems,
but also the separation
between the quantum and classical physics in general.
Plus, last but not least, of the
further desirable contribution to the search for
mutual parallels between the Schr\"{o}dinger equation
for quantum systems and its very close formal
parallel of Maxwell equations in paraxial approximation
in photonics.}

\textcolor{black}{In the latter setting of photonics,
the mathematics using non-Hermitian generators of evolution
which was originally developed in quantum-physics framework
found a truly productive domain of
innovative applications \cite{Christodoulides}.
Many of the authors
paid also an exquisite attention
to the study of
the evolution dynamics
near the exceptional-point regime.
Having all of these practical needs in mind,
we still felt forced to narrow the scope of the present paper.
Using
the  perturbation-motivated
replacement of
a given $g-$dependent Hamiltonian $H(g)$
by its avatar
 $
  P(\lambda)$,
we decided to analyze, first of all, the
problem of the smallness of
the perturbations in $P(\lambda)$.
Due to the domination of this matrix
by an unperturbed
(as well as unphysical) Jordan-matrix limit,
we decided to put main emphasis
upon the fact that what seems to be most desirable is
an amendment of
insight
in the
perturbative
solutions.}

\textcolor{black}{With a broader context being clarified, let us now return again to the
narrower research field in which the spectrum is required real.}
According to the general analysis of Eq.~(\ref{zucose})
as performed in \cite{pertEP}
we know that near
the EP4 singularity,
the reality or non-reality
of the spectrum
is only influenced by the matrix elements of
$P^{(4)}(\lambda)$ which are
localized below the main diagonal.
Without any loss of generality, therefore,
it is sufficient to study,
for our present purposes (of
a guarantee of the reality of the spectrum near EP4),
just the following six-parametric
and properly scaled perturbed-Jordan-matrix
Hamiltonian
 \be
 \label{ugb}
 P^{(4)}_{(a,b,c,x,y,z)}(\lambda)=
 \left[ \begin {array}{cccc} 0&1&0&0
 \\\noalign{\medskip}{{\lambda}}^{2}z&0&1&0
 \\\noalign{\medskip}{{\lambda}}^{3}x&{{\lambda}}^{2}y&0&1
 \\\noalign{\medskip}{{\lambda}}^{4}
 a &{{\lambda}}^{3}
 b &{{\lambda}}^{2} c &0
 \end {array} \right]
 %=P^{(4)}(\lambda,a,b,c,x,y,z)
 \,.
 \ee
We are now going to demonstrate that
such a reduced-matrix subfamily of
our canonical
non-Hermitian Hamiltonians
can be interpreted as a source of an exhaustive
qualitative
classification of spectra of the
phenomenological input-information-carrying
Hamiltonians $H^{(4)}(g)$, connected with
its canonical isospectral avatar of Eq.~(\ref{ugb})
by
the transition-matrix-mediated
correspondence of Eq.~(\ref{pertie}).

%\newpage

\section{Conditions of unitarity\label{segma}}

Under
the above-mentioned methodical guidance by paper \cite{pertEP}
it makes sense to
reparametrize the eigenvalues $\eta=\eta({\lambda})$ of $H^{(4)}(g)$
{\it alias\,}
$P^{(4)}({\lambda})$
of Schr\"{o}dinger Eq.~(\ref{zucose})
as follows,
 $$
 \eta({\lambda})={\lambda}\,E({\lambda})\,.
 $$
We are now prepared to prove the existence of a unitarity-compatible
(i.e., of the reality-of-spectrum supporting
{\it alias\,} stable-bound-states-supporting)
vicinity ${\cal D}_{physical}$ of the
EP4 singularity in the space of parameters.

\subsection{Secular equation}

The real four-by-four-matrix form (\ref{ugb})
of our perturbed canonical Hamiltonian
is asymmetric. During a decrease of the
absolute value of
parameter $\lambda$ to zero,
this parameter  measures
the smallness
of perturbation corrections.
It is not too surprising that
this quantity enters the Hamiltonian matrix
in a nonlinear manner which
weights
the influence of its
individual matrix elements, ordering them
as proportional to their distance from the main diagonal.

Most easily, the latter consequence of the general theory
can be very easily verified
when we write down our $N=4$ secular equation
determining the
quadruplet of the energies $\eta=\lambda\,E$,
 \be
 \label{rugby}
 S(E)=
 \det
 \left[ \begin {array}{cccc} -{\lambda}\,E&1&0&0
 \\\noalign{\medskip}{{\lambda}}^{2}z&-{\lambda}\,E&1&0
 \\\noalign{\medskip}{{\lambda}}^{3}x&{{\lambda}}^{2}y&-{\lambda}\,E&1
 \\\noalign{\medskip}{{\lambda}}^{4}
 a &{{\lambda}}^{3}
 b &{{\lambda}}^{2} c &-{\lambda}\,E
 \end {array} \right]=0\,.
 \ee
The construction
of the solutions becomes simplified
when we reparametrize the lowest row of the
elements of the perturbed Hamiltonian matrix as follows,
 $$
  c= c(\gamma)=\gamma-y-z
 \,,\ \ \
 b= b(\beta)=\beta-x
\,\ \ \
 a= a(\alpha)=\alpha-z \left( y-\gamma \right) -{z}^{2}
 \,.
 $$
This leads to a maximally compact four-parametric characteristic polynomial
 $$
 {{\it {\eta}}}^{4}-{{\lambda}}^{2}\gamma\,{{\it {\eta}}}^{2}-{{\lambda}}^{3}\beta\,
 {\it {\eta}}-{{\lambda} }^{4}\alpha
 $$
which
specifies
the complete set of the three
${\lambda}-$dependent energy roots ${\eta}={\eta}({\lambda})$, real or complex.
Alternatively,
in terms of a rescaled energy $E$
we obtain the simpler, three-parametric secular equation
 \be
 S(E)= {{\it {E}}}^{4}-\gamma\,
 {{\it {E}}}^{2}-\beta\,{\it {E}}-\alpha=0\,.
 \label{ure}
 \ee
It determines
the quadruplet of the rescaled-energy roots
$E=E_n(\alpha,\beta,\gamma)$,
$n=0,1,2,3$
which may be real or complex, and
which remain all ${\lambda}-$independent.
Thus, the experimental bound-state
or resonant-state energies $\eta_n({\lambda})$
will be the linear functions of
parameter ${\lambda}$. Near the EP4 degeneracy
this
yields the unfolding perturbed
energies, real or complex, in explicit form.

\subsection{Corridor of \textcolor{black}{unitary} access to the degeneracy}

As we repeatedly emphasized,
our interest in the dynamics of quantum
systems near
their EP4 degeneracies
is mainly restricted to
the quasi-Hermitian
systems which satisfy constraint (\ref{[7]})
and which possess a strictly real energy spectrum.
As long as such a spectrum is given by
the roots of secular equation $S(E)=0$,
we only have to
recall its explicit three-parametric $N=4$
form (\ref{ure}). On these grounds,
we finally have to determine the boundaries of
the physical domain ${\cal D}_{}$.

\subsubsection{Construction}

The requirement
of the unitarity of the evolution
in the perturbed dynamical regime with
physical ${\lambda} \neq 0$
will only be satisfied
for the triplet of parameters $\alpha=\alpha_{physical}$,
$\beta=\beta_{physical}$ and $\gamma=\gamma_{physical}$
belonging to a real and open
three-dimensional physical
domain,
 \be
 \{\alpha_{physical},\beta_{physical},\gamma_{physical}\} \in
 {\cal D}={\cal D}_{physical}\,.
 \label{zereal}
 \ee
The elementary form of the secular Eq.~(\ref{ure}) will now be used
to localize the boundaries $\partial {\cal D}$. This means that
inside ${\cal D}$ we may require the non-degeneracy and reality of
the quadruplet of the bound-state energy roots. Let us denote and
order these numbers as follows,
 \be
 E^{[4]}_{- -} < E^{[4]}_{-} < E^{[4]}_{+} < E^{[4]}_{++}\,.
 \label{hlav}
 \ee
%
%\subsection{The proof of the reality of the perturbed spectrum}
%
In the light of such a requirement
%(\ref{hlav})
the graph of our secular polynomial $S(E)$ of Eq.~(\ref{ure})
has to have the two
negative minima
(say, at an ordered pair of certain real values of $E=E^{[3]}_{\pm}$
to be specified later) and one positive maximum
(in the middle, at another real $E=E^{[3]}_{0}$).
At these extremes, therefore, it is necessary and sufficient to have
 \be
 S(E^{[3]}_{-})<0\,,\ \ \
 S(E^{[3]}_{0})>0\,,\ \ \
 S(E^{[3]}_{+})<0\,.
 \label{reconstr4}
 \ee
The ordered triplet
 \be
 E^{[3]}_{-} < E^{[3]}_{0} < E^{[3]}_{+}
 \label{constr3}
 \ee
of the
coordinates of these extremes
coincides
with the roots of the third-order polynomial
 \be
 S'(E)= 4\,{\it {E}}^{3}-2\,\gamma\,{{\it {E}}}^{}-\beta\,.
 \label{ure1}
 \ee
\textcolor{black}{The mutual relationship between
the local extremes (\ref{reconstr4})
of the secular
polynomial $S(x)$ of Eq.~(\ref{ure})
and the zeros (\ref{constr3})
of its derivative $S'(x)$ of Eq.~(\ref{ure1})
is a very core of our present considerations.
For its explicit illustration
let us pick up
a sample set of parameters
 \be
 \{\alpha_{},\beta_{},\gamma_{}\}
 =
 \{-24,-10,15\}
 \label{kozereal}
 \ee
which
yield
the specific secular polynomial
$S(x)$,
and which have to be tested whether they truly
lie safely inside ${\cal D}_{physical}$.
As long as both the secular polynomial
$S(x)={x}^{4}- 15\,{x}^{2}+ 10\,x+ 24$ and its derivative
$S'(x)=4\,{x}^{3}- 30\,{x}^{}+ 10$
are extremely elementary,
the simplest version of the test
can be just numerical and graphical.
Its affirmative result
can immediately be read off
Figure \ref{Qws5} in which
we
see that all of the four roots of $S(x)$
(marked, in the picture, by the upper quadruplet of small circles)
are real,
and that its maxima
really occur at the roots of the other polynomial $S'(x)$
(for the sake of the clarity of the picture,
both of the graphs are mutually shifted and rescaled,
i.e., displayed in arbitrary units).}

\textcolor{black}{
\begin{figure}[t]                    %instead of \begin{figure}[t]
                   %instead of \begin{center}
\epsfig{file=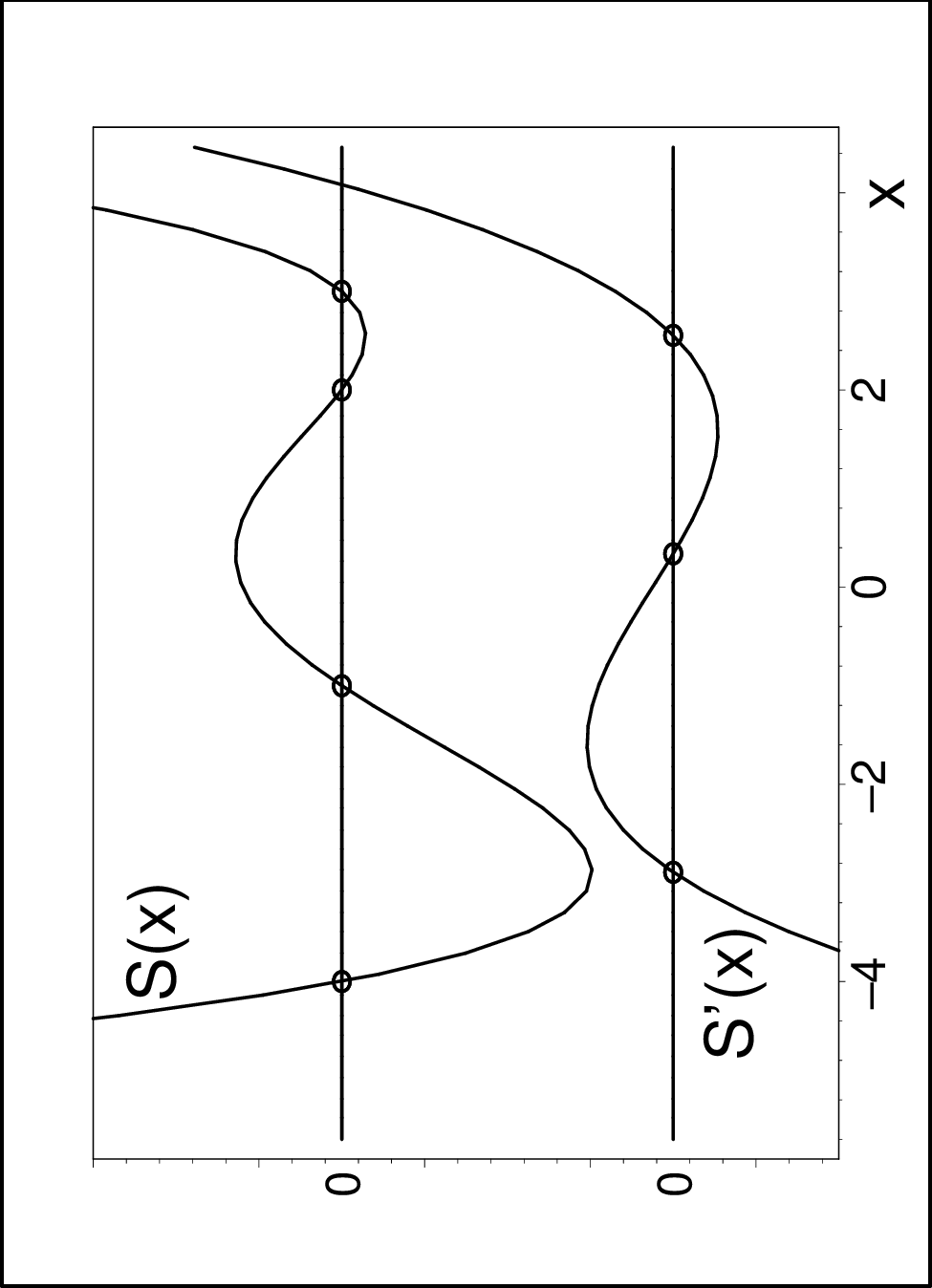,angle=270,width=0.35\textwidth}
   % \sidecaption                      %instead of \end{center}
%\vspace{2mm}
\caption{The roots of the toy-model $S(x)$ (small circles in the upper graph)
and of $S'(x)$ (small circles in the shifted, lower graph).}
 \label{Qws5}
\end{figure}
}

%nasleduje

\textcolor{black}{During the subsequent continued rigorous analysis, an}
analogous argumentation can be applied: The graph of function
$S'(E)$ has to have one maximum (say, at some smaller real value of
$E=E^{[2]}_{-}$) and one minimum (at a larger real value of
$E=E^{[2]}_{+}$). At these extremes it is necessary and sufficient
to have
 \be
 S'(E^{[2]}_{-})>0\,,\ \ \
 S'(E^{[2]}_{+})<0\,.
 \label{condy}
 \ee
In a complete analogy with the preceding step, the
coordinates of the latter extremes
are easily identified with the roots of polynomial
 \be
 S''(E)= 12\,{\it {E}}^{2}-2\,\gamma\,.
 \label{ure2}
 \ee
Thus,
the necessary and sufficient reality of these roots
can be easily guaranteed
via a trivial reparametrization (i.e., via the positivity) of
 \be
 \gamma_{physical}=6\,\kappa^2\,,\ \ \ \ \kappa>0\,.
 \label{posi}
 \ee
This yields the closed formula for
 $
 E^{[2]}_{\mp} =\mp \kappa\,.
 %\label{reconstr2}
 $
and, ultimately, this also converts the pair of inequalities
(\ref{condy})
into the following
explicit constraint imposed upon the
admissible physical
values of parameter $\beta$,
 \be
  -8\,\kappa^3< \beta_{physical} <8\,\kappa^3\,.
  \label{inte}
 \ee
What remains for us to guarantee is the validity
of relations (\ref{reconstr4}).

This is the task which requires the
evaluation of
the triplet of the roots (\ref{constr3})
of the cubic polynomial (\ref{ure1}).
The
strictly algebraic and exhaustive exact
answer
(i.e., an explicit specification of the admissible range
of $\alpha_{physical}$)
can be provided in terms of
the well known Cardano formulae.
For our present purposes, nevertheless,
we intend to skip the display of these formulae.
In order to provide a more intuitive insight in the
structure of the boundary of the physical domain ${\cal D}$,
we will use a less formal alternative approach.

\subsubsection{EP4-unfolding paths with vanishing $\beta$}

In order to simplify the analysis, let us first
pick up a special, sample value of
 $$
 \beta=\beta_{physical}^{(sample)}=0\,.
 $$
This will enable us to deduce the related,
most elementary sample
forms of
 \be
 E^{[3](sample)}_{\mp}=\mp \sqrt{3}\,\kappa\,,\ \ \ \
 E^{[3](sample)}_{0}=0\,,
 \label{samstr3}
 \ee
i.e., of the coordinates to be inserted in Eq.~(\ref{reconstr4}).
From its middle item we get the upper bound
while the other two inequality items Eq.~(\ref{reconstr4})
give the same lower-bound constraint. Thus, we arrive at the
following special, $\beta=0$ boundaries,
 \be
  0>\alpha_{physical}^{(sample)}>-9\,\kappa^4\,.
  \label{lowerb}
 \ee
The
domain is fully determined.
In fact, the most important consequence of this result is that
it implies that in a sufficiently small vicinity of the EP4 degeneracy,
the
physical quasi-Hermiticty domain ${\cal D}$ is not empty because
even its ``central'', $\beta=0$
subdomain
${\cal D}_{physical}^{(sample)}$ is non-empty.
There exists a unitary-evolution
path of possible fall of the system
into an EP4-mediated quantum phase transition.

\subsubsection{Small but non-vanishing choice of $\beta$}

After a small change (i.e., say, growth)
of
 \be
  \beta=\beta_{physical}^{(perturbed)} =\delta^2\,\kappa^3\,,
  \ \ \ \ \delta \ll 1
  \label{asss}
  \ee
we get, at the small $\delta$s,
 $$
  %E^{[3](perturbed)}_{-}\approx -\kappa\,
% \sqrt{3-\sqrt{3}\,
%  \delta^2/12}\,,
%  \ \ \ \ \
  E^{[3](perturbed)}_{0}\approx -\kappa\,\delta^2/12\,,
  \ \ \ \ \
  E^{[3](perturbed)}_{\mp }\approx \mp \kappa\,
 \sqrt{3\mp \sqrt{3}\,
  \delta^2/12 }\,,
 $$
i.e.,
we detect the  decrease of $E^{[3](perturbed)}_{0}$ and the
growth of $E^{[3](perturbed)}_{-}$
(plus
the
growth of $E^{[3](perturbed)}_{+}$
which is, due to our assumption (\ref{asss}), irrelevant).
This implies
the growth of the lower bound for
   %$\alpha_{physical}^{(perturbed)}$,
 $$
 \alpha_{physical}^{(perturbed)}/\kappa^4 \gtrapprox -9 +\sqrt{3}\,\delta^2
 $$
Also the upper bound of the same quantity becomes modified
by perturbation (\ref{asss}),
 $$
 \alpha_{physical}^{(perturbed)}/\kappa^4 \lessapprox \delta^4/24\,.
 $$
What can be deduced is the fact that the
interval of
the admissible $\alpha$s remains non-empty, and that it
shrinks and moves to the right.

\subsubsection{The limit of a maximal size of $\beta$}

Out of the two
maximal-size
choices of $\beta$
(representing, strictly speaking,
just an auxiliary, limiting,
EP-degenerate case)
let us pick up, for illustration, just the positive one,
 $$
  \beta=\beta_{physical}^{(extreme)} =8\,\kappa^3\,.
  %\label{inte}
 $$
This choice implies
the following values of the ultimate bounds imposed upon the
following triplet of roots,
 \be
 E^{[3](extreme)}_{-}=
 E^{[3](extreme)}_{0}=-\kappa\,,\ \ \ \\ \ \ \
 E^{[3](extreme)}_{+}=2\,\kappa\,.
 \label{exsamstr3}
 \ee
Finally we obtain the following unique limiting value of
 $$
 \alpha=\alpha_{physical}^{(extreme)}=3\,\kappa^4\,,
 $$
confirming both of the above-indicated tendencies of the
movement
of the interval of the admissible $\alpha$s
to the right. The process of its shrinking
ends up with the
ultimate zero length of the interval,
and its position
is found to have moved
to the right from zero,
yielding the ultimate positive values
of all of the eligible $\alpha$s in this extreme.

%\newpage

\section{Discussion\label{discussion}}

Half a century ago, the notion of exceptional points
only served the needs of the
perturbation theory of linear operators \cite{Kato},
but it did not last too long before it also
found multiple other applications
in quantum physics \cite{BW,Alvarez,BB},
especially in the context of the so called open
and/or resonant quantum systems
\cite{Nimrod}.
Moreover,
the notion of EPs also appeared
ubiquitous in
several other
areas of physics
including even the non-quantum ones
\cite{Rotter,Mousse,Christodoulides,Berry,Heiss,Nimroda}.

Several open theoretical questions
are currently emerging
in this setting.
A remarkable
progress
in their understanding
has been
achieved,
in particular, after people recognized that
the unitarity of the evolution
(considered,
traditionally, robust
and guaranteed by the Stone theorem
by which the unitarity of the evolution follows from the ``obligatory''
Hermiticity of the Hamiltonian \cite{Stone})
is in fact conditional.
It has been revealed, indeed, that
it may make sense to
work with more Hilbert spaces
or, in some cases, even with just two or more
eligible inner products \cite{hybrid}.

Currently,
it is widely accepted that
the traditional requirement of
the Hermiticity of the operators of observables
can be replaced by a
less restrictive mathematical requirement of their
quasi-Hermiticity {\it alias\,}
Hermitizability \cite{book}.
Several formal advantages of such an amendment
of the formalism of quantum mechanics
(cf., e.g., its
older comprehensive reviews \cite{Geyer,Carl,ali})
are well recognized at present: They can even be
well sampled in
the
Buslaev's and Grecchi's fairly old
study of
certain quantum anharmonic oscillators \cite{BG}, etc.

One of the decisive practical
advantages of working with the non-Hermitian but
Hermitizable ({\it alias\,} quasi-Hermitian)
operators and Hamiltonians
can be seen, in similar examples,
in the possibility
of an optimal
representation of a quantum system
in two
(or in three \cite{SIGMA}, or even in more than three \cite{more})
different Hilbert spaces.
The information about dynamics
as carried, traditionally, just by the Hamiltonian $H(g)$
is then shared and distributed between the Hamiltonian
and another operator  $\Theta$
called
physical inner-product metric.

One of the most important consequences of the
new flexibility introduced via
the unusual $\Theta \neq I$
appeared to make multiple quantum models
conceptually closer to its
classical partners.
Such an enhancement of the contacts with classical physics
also inspired
new forms of an insight in
the genuine quantum structures.
In our present paper we addressed,
in this sense,
the emergence of singularities
which found, in their classical forms,
their appropriate mathematical description
in the language of geometry: The popular Thom's
catastrophe theory \cite{Zeeman}
could be recalled as one of the
fairly universal
mathematical formalisms
of such a type.

In the context of quantum theory the
latter form of inspiration is of a comparatively new date.
One of the reasons is that the
study of the quantum singular behavior
is much more complicated, for several reasons
ranging from the underlying Hilbert-space-based
theory up to the probabilistic nature of
its experimental tests.
Fortunately, the recent growth of popularity
of certain non-standard forms of the
Hilbert space (in which, say, the operator representing a
quantum  observable need not be self-adjoint
\cite{book,Geyer,Dyson,Carl,ali})
opened also a number of innovative
approaches to the formulation of the
concept of a quantum singularity.

Some of the related and most successful models of a
singularity in a quantum system
(involving, typically, a quantum phase transition \cite{passage})
were based on the use of the Kato's \cite{Kato} concept of
exceptional point $g^{(EP)}$.
In this sense,
the results described in our present paper belong
precisely to the
latter context.
With our project being inspired by
certain
technical obstacles as encountered in
the literature
(and,
in particular, in
paper \cite{Bijan}),
we imagined that
precisely due to
the existence of
these technical obstacles,
the progress in the related applied physics
is comparatively slow:
Typically, most authors
were circumventing the technical challenges and
feelt forced to
restrict their
analyses of quantum phase transitions
to the EPN models with $N=2$ or $N=3$.
This was, in fact, the main reason why we
redirected here the detailed technical analysis
to the
general class of quantum systems with $N=4$.

%------------------------------------------------------------------

%\newpage

%\subsection*{Funding:}
%
%The project did not receive any support.

\subsection*{Data availability statement:}

No new data were created or analyzed in this study.

\subsection*{Conflicts of Interest:}

The author declares no conflicts of interests.

%%\newpage
\subsection*{ORCID ID:}

Miloslav Znojil: https://orcid.org/0000-0001-6076-0093

\end{document}